\documentclass[letter]{aa}
\usepackage{natbib}
\usepackage{graphicx}
\usepackage{txfonts}

\begin{document}

\title{GRB060218 and GRBs associated with Supernovae Ib/c}

\author{
Maria Giovanna Dainotti\inst{1,2}
\and
Maria Grazia Bernardini\inst{1,2}
\and
Carlo Luciano Bianco\inst{1,2}
\and
Letizia Caito\inst{1,2}
\and
Roberto Guida\inst{1,2}
\and
Remo Ruffini\inst{1,2,3}
}

\institute{
ICRANet and ICRA, Piazzale della Repubblica 10, I-65122 Pescara, Italy.
\and
Dipartimento di Fisica, Universit\`a di Roma ``La Sapienza'', Piazzale Aldo Moro 5, I-00185 Roma, Italy. E-mails: dainotti@icra.it, maria.bernardini@icra.it, bianco@icra.it, letizia.caito@icra.it, roberto.guida@icra.it, ruffini@icra.it.
\and
ICRANet, Universit\'e de Nice Sophia Antipolis, Grand Ch\^ateau, BP 2135, 28, avenue de Valrose, 06103 NICE CEDEX 2, France.
}

\titlerunning{GRB060218 and GRBs associated with Supernovae Ib/c}

\authorrunning{Dainotti et al.}

\date{}

\abstract{
The \emph{Swift} satellite has given continuous data in the range $0.3$--$150$ keV from $0$ s to $10^6$ s for GRB060218 associated with SN2006aj. This GRB which has an unusually long duration ($T_{90}\sim 2100$ s) fulfills the Amati relation. These data offer the opportunity to probe theoretical models for Gamma-Ray Bursts (GRBs) connected with Supernovae (SNe).
}{
We plan to fit the complete $\gamma$- and X-ray light curves of this long duration GRB, including the prompt emission, in order to clarify the nature of the progenitors and the astrophysical scenario of the class of GRBs associated to SNe Ib/c. 
}{
We apply our ``fireshell'' model based on the formation of a black hole, giving the relevant references. It is characterized by the precise equations of motion and equitemporal surfaces and by the role of thermal emission.
}{
The initial total energy of the electron-positron plasma $E_{e^\pm}^{tot} = 2.32\times 10^{50}$ erg has a particularly low value similarly to the other GRBs associated with SNe. For the first time we observe a baryon loading $B =10^{-2}$ which coincides with the upper limit for the dynamical stability of the fireshell. The effective CircumBurst Medium (CBM) density shows a radial dependence $n_{cbm} \propto r^{-\alpha}$ with $1.0 \la \alpha \la 1.7$ and monotonically decreases from $1$ to $10^{-6}$ particles/cm$^3$. Such a behavior is interpreted as due to a fragmentation in the fireshell. Analogies with the fragmented density and filling factor characterizing Novae are outlined. The fit presented is particularly significant in view of the complete data set available for GRB060218 and of the fact that it fulfills the Amati relation. 
}{
We fit GRB060218, usually considered as an X-Ray Flash (XRF), as a ``canonical GRB'' within our theoretical model. The smallest possible black hole, formed by the gravitational collapse of a neutron star in a binary system, is consistent with the especially low energetics of the class of GRBs associated with SNe Ib/c. We give the first evidence for a fragmentation in the fireshell. Such a fragmentation is crucial in explaining both the unusually large $T_{90}$ and the consequently inferred abnormal low value of the CBM effective density. 
}

\keywords{gamma rays: bursts --- black hole physics --- (stars:) binaries: general --- stars: neutron}

\maketitle

\section{Introduction}

GRB060218, discovered by the \emph{Swift} satellite \citep{ma06} with cosmological redshift $z=0.033$ \citep{Mi06,Sollermann06}, is the best example of a Gamma-Ray Burst (GRB) associated with a Supernova (SN) Ib/c \citep{caa06}. Its extremely long duration is peculiar, with the longest $T_{90}$ ever observed ($T_{90}\sim 2100$ s). $T_{90}$ is defined as the time over which a burst emits from $5$\% to $95$\% of its total measured energy in the prompt emission. This definition depends however on the instrumental threshold \citep[see][for details]{Venezia_Orale}. This source is also interesting since it represents a discriminant between existing GRB theories: it has been pointed out by \citet{So06b} and \citet{Fan06} that it is impossible to explain the X- and radio afterglows within the traditional synchrotron model. They attempted to fit only the late radio data after $\sim 10^3$ s and they attributed the nature of the prompt emission to a yet undetermined ``inner engine'' \citep[see][]{So06b}, possibly a magnetar \citep{Maz06}.

In this Letter we present a detailed fit of the entire X- and $\gamma$-ray light curves including the prompt emission: there is no need here for the prolonged activity of an inner engine. Therefore we explain the unusually high values of the observed $T_{90}$ by our ``fireshell'' model \citep[see sec. \ref{sec1},][and references therein]{rlet1,rlet2,rubr,rubr2,EQTS_ApJL,EQTS_ApJL2,PowerLaws}.

After summarizing our model in sec. \ref{sec1}, in sec. \ref{sec2} we recall GRB060218's observational data. In sec. \ref{fit} we show the fit of the BAT and XRT light curves (in the $15$--$150$ keV and in the $0.3$--$10.0$ keV energy bands respectively, see Figs. \ref{tot},\ref{global2}). In Fig. \ref{global} and sec. \ref{sec4} we discuss the actual and effective CircumBurst Medium (CBM) density. We outline the occurrence of a fragmentation in the fireshell pointing out some analogies with the ejecta of Novae. We then proceed to the general conclusions.

\section{The fireshell model}\label{sec1}

We assume that all GRBs, whether ``short'' or ``long'', originate from the gravitational collapse to a black hole \citep{rlet2}. The $e^\pm$ plasma created in the process of the black hole formation expands as a spherically symmetric ``fireshell'' with a constant width in the laboratory frame, i.e. the frame in which the black hole is at rest. We have only two free parameters characterizing the source, namely the total energy $E_{e^\pm}^{tot}$ of the $e^\pm$ plasma and its baryon loading $B\equiv M_Bc^2/E_{e^\pm}^{tot}$, where $M_B$ is the total baryons' mass \citep{rswx00}. They fully determine the optically thick acceleration phase of the fireshell, which lasts until the transparency condition is reached and the Proper-GRB \citep[P-GRB, see][]{rlet2} is emitted. Then, the afterglow emission starts due to the collision between the remaining optically thin fireshell and the CBM, and it clearly depends on the parameters describing the effective CBM distribution (see below). The luminosity of such an afterglow emission consists of a rising branch, a peak, and a decaying tail \citep{rlet2}.

Therefore, unlike treatments in the current literature \citep[see e.g.][and references therein]{p04}, in our model we define a ``canonical GRB'' light curve with two sharply different components: the P-GRB and the afterglow \citep{rlet2,XIIBSGC}. The ratio between the total energies of these two components and the temporal separation between their peaks are functions of the $B$ parameter \citep{rlet2}. The peak of the afterglow contributes to what is usually called the GRB ``prompt emission'' \citep[see e.g.][]{rlet2,050315}.

Another crucial assumption is that the afterglow luminosity is due to a thermal emission in the co-moving frame of the fireshell \citep{spectr1}. The ${\cal R}$ parameter defines the temperature $T$ of such a thermal emission:
\begin{equation}
{\cal R} \equiv \frac{A_{eff}}{A_{vis}} = \frac{dE/dt}{4\pi r^2 \sigma T^4}\, ,
\label{Rdef}
\end{equation}
where $dE/dt$ is the source luminosity, $\sigma$ is the Stefan-Boltzmann constant, $r$ is the radius of the fireshell, $A_{eff}$ is its effective emitting area and $A_{vis}$ is its total visible area. ${\cal R}$ and the CBM effective density $n_{cbm}$ are the two parameters which fully describe the effective CBM distribution taking into account its filamentary structure \citep{fil}. Similar considerations in a different context has been recently presented in \citet{pa07}.

The description of the engulfment of the CBM matter by the fireshell is a most complex and time consuming procedure. In the non-relativistic systems such a description can be made at each point. In this ultrarelativistic regime a more global approach is needed. The arrival time of each photon at the detector depends on the entire \textit{past} history of the fireshell \citep{rlet1}. All the observables depends on the equitemporal surfaces \citep[EQTSs,][]{EQTS_ApJL,EQTS_ApJL2} which produce observable effects in a large part of the \textit{future} history of the fireshell. The CBM structure due to its feedback on the equation of motion of the fireshell must be therefore inferred self-consistently with this very non-linear evolution of the entire fireshell history. These difficulties are further increased by the necessity to fit the observed light curves in selected energy bands \citep[two in the present case of GRB060218 and up to five in the case of GRB050315, see][]{050315}. The fulfillment of these constraints represents a severe test not only for the validity of the theory but also for the spectral models assumed in the data reduction.

The fireshell model has been successfully applied to GRB050315 \citep{050315}, GRB031203 \citep{031203}, GRB980425 \citep{Mosca_Orale}, GRB030329 \citep{030329}, GRB970228 (Bernardini et al., in preparation), GRB991216 \citep{rubr,spectr1}. Not all these sources fulfills the correlation proposed by \citet{aa02} between the isotropic equivalent energy emitted in the prompt emission and the peak energy of the corresponding time-integrated spectrum (see Dainotti et al., in preparation).

\section{GRB060218 - SN2006aj}\label{sec2}

GRB060218 triggered the BAT instrument of {\em Swift} on 18 February 2006 at 03:36:02 UT and has a $T_{90} = (2100 \pm 100)$ s \citep{ca06}. The XRT instrument \citep{Ka06,ca06} began observations $\sim 153$ s after the BAT trigger and continued for $\sim 12.3$ days \citep{Sa06}. The source is characterized by a flat $\gamma$-ray light curve and a soft spectrum \citep{Ba06}. It has an X-ray light curve with a long, slow rise and gradual decline and it is considered an X-Ray Flash (XRF) since its peak energy occurs at $E_p=4.9^{+0.4}_{-0.3}$ keV \citep{caa06}. It has been observed by the \emph{Chandra} satellite on February 26.78 and March 7.55 UT ($t\simeq 8.8$ and $17.4$ days) for $20$ and $30$ ks respectively \citep{So06b}. The spectroscopic redshift has been found to be $z=0.033$ \citep{Sollermann06,Mi06}. The corresponding isotropic equivalent energy is $E_{iso}=(1.9\pm 0.1)\times 10^{49}$ erg \citep{Sa06} which sets this GRB as a low luminous one, consistent with most of the GRBs associated with SNe \citep{Liang06,Cob06,Gue06}.

GRB060218 is associated with SN2006aj whose expansion velocity is $v\sim 0.1c$ \citep{Pi06,fa06,So06a,Cob06}. The host galaxy of SN2006aj is a low luminosity, metal poor star forming dwarf galaxy \citep{Fe06} with an irregular morphology \citep{Wi07}, similar to the ones of other GRBs associated with SNe \citep{Modjaz06,Sollermann06}.

\section{The fit of the observed data}\label{fit}

\begin{figure}
\includegraphics[width=\hsize,clip]{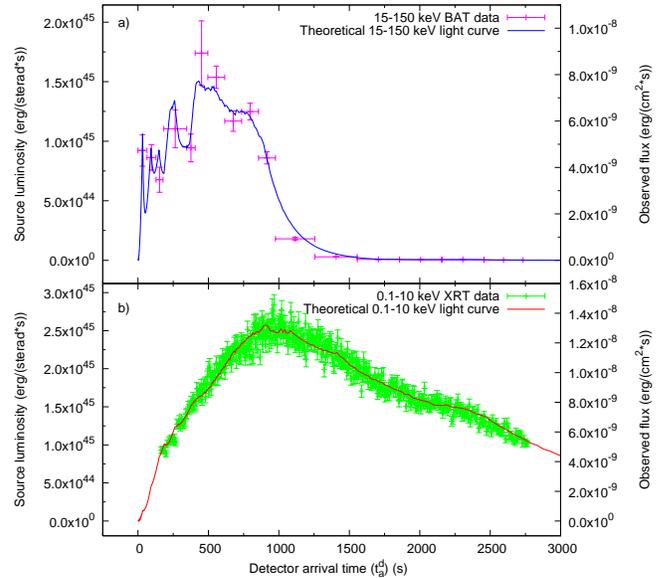}
\caption{GRB060218 prompt emission: a) our theoretical fit (blue line) of the BAT observations in the $15$--$150$ keV energy band (pink points); b) our theoretical fit (red line) of the XRT observations in the $0.3$--$10$ keV energy band (green points) \citep[Data from:][]{caa06}.}
\label{tot}
\end{figure}

\begin{figure}
\includegraphics[width=\hsize,clip]{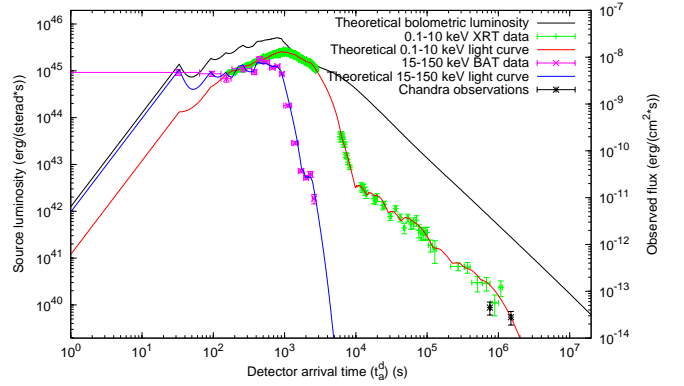}
\caption{GRB060218 complete light curves: our theoretical fit (blue line) of the $15$--$150$ keV BAT observations (pink points), our theoretical fit (red line) of the $0.3$--$10$ keV XRT observations (green points) and the $0.3$--$10$ keV \textit{Chandra} observations (black points) are represented together with our theoretically computed bolometric luminosity (black line) \citep[Data from:][]{caa06,So06b}.}
\label{global2}
\end{figure}

In this section we present the fit of our fireshell model to the observed data (see Figs. \ref{tot}, \ref{global2}). The fit leads to a total energy of the $e^\pm$ plasma $E_{e^\pm}^{tot}= 2.32\times 10^{50}$ erg, with an initial temperature $T = 1.86$ MeV and a total number of pairs $N_{e^\pm} = 1.79\times 10^{55}$. The second parameter of the theory, $B = 1.0 \times 10^{-2}$, is the highest value ever observed and is close to the limit for the stability of the adiabatic optically thick acceleration phase of the fireshell \citep[for further details see][]{rswx00}. The Lorentz gamma factor obtained solving the fireshell equations of motion \citep{EQTS_ApJL2,PowerLaws} is $\gamma_\circ=99.2$ at the beginning of the afterglow phase at a distance from the progenitor $r_\circ=7.82\times 10^{12}$ cm. It is much larger than $\gamma \sim 5$ estimated by \citet{Kan06} and \citet{Toma06}.

In Fig. \ref{tot} we show the afterglow light curves fitting the prompt emission both in the BAT ($15$--$150$ keV) and in the XRT ($0.3$--$10$ keV) energy ranges, as expected in our ``canonical GRB'' scenario (see Dainotti et al., in preparation). Initially the two luminosities are comparable to each other, but for a detector arrival time $t_a^d > 1000$ s the XRT curves becomes dominant. The displacement between the peaks of these two light curves leads to a theoretically estimated spectral lag greater than $500$ s in perfect agreement with the observations \citep[see][]{la06}. We obtain that the bolometric luminosity in this early part coincides with the sum of the BAT and XRT light curves (see Fig. \ref{global2}) and the luminosity in the other energy ranges is negligible.

We recall that at $t_a^d \sim 10^4$ s there is a sudden enhancement in the radio luminosity and there is an optical luminosity dominated by the SN2006aj emission \citep[see][]{caa06,So06b,Fan06}. Although our analysis addresses only the BAT and XRT observations, for $r > 10^{18}$ cm corresponding to $t_a^d > 10^4$ s the fit of the XRT data implies two new features: \textbf{1)} a sudden increase of the ${\cal R}$ factor from ${\cal R} = 1.0\times 10^{-11}$ to ${\cal R} = 1.6\times 10^{-6}$, corresponding to a significantly more homogeneous effective CBM distribution (see Fig.\ref{global}b); \textbf{2)} an XRT luminosity much smaller than the bolometric one (see Fig. \ref{global2}). These theoretical predictions may account for the energetics of the enhancement of the radio and possibly optical and UV luminosities. Therefore, we identify two different regimes in the afterglow, one for $t_a^d < 10^4$ s and the other for $t_a^d > 10^4$ s. Nevertheless, there is a unifying feature: the determined effective CBM density decreases with the distance $r$ monotonically and continuously through both these two regimes from $n_{cbm} = 1$ particle/cm$^3$ at $r = r_\circ$ to $n_{cbm} = 10^{-6}$ particle/cm$^3$ at $r = 6.0 \times 10^{18}$ cm: $n_{cbm} \propto r^{-\alpha}$, with $1.0 \la \alpha \la 1.7$ (see Fig. \ref{global}a).

Our assumption of spherical symmetry is supported by the observations which set for GRB060218 an opening beaming angle larger than $\sim 37^\circ$ \citep{Liang06,caa06,So06b,Gue06}.

\section{The fireshell fragmentation}\label{sec4}

\begin{figure}
\includegraphics[width=\hsize,clip]{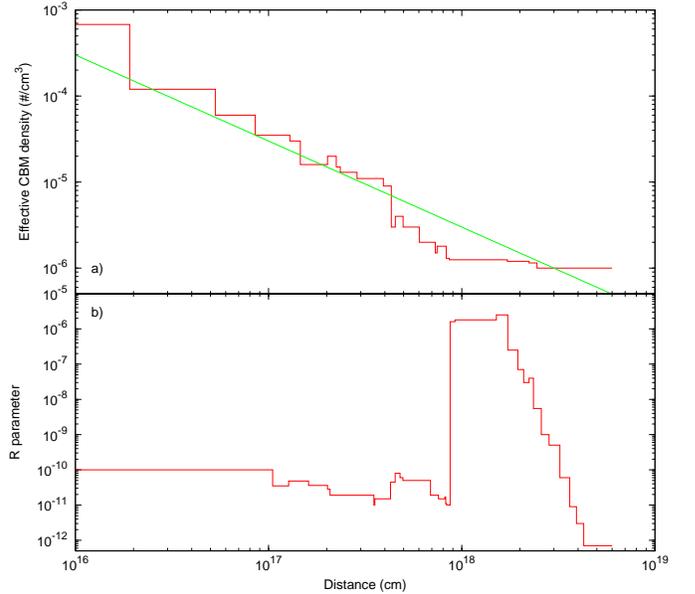}
\caption{The CBM distribution parameters: a) the effective CBM number density (red line) monotonically decreases with the distance $r$ following Eq.(\ref{nismact}) (green line); b) the ${\cal R}$ parameter vs. distance.}
\label{global}
\end{figure}

GRB060218 presents different peculiarities: the extremely long $T_{90}$, the very low effective CBM density decreasing with the distance and the largest possible value of $B=10^{-2}$. These peculiarities appear to be correlated. Following \citet{Mosca_Orale}, we propose that in the present case the fireshell is fragmented. This implies that the surface of the fireshell does not increase any longer as $r^2$ but as $r^\beta$ with $\beta < 2$. Consequently, the effective CBM density $n_{cbm}$ is linked to the actual one $n_{cbm}^{act}$ by:
\begin{equation}
n_{cbm} = {\cal R}_{shell} n_{cbm}^{act}\, , \quad \mathrm{with} \quad {\cal R}_{shell} \equiv \left(r^\star/r\right)^\alpha\, ,
\label{nismact}
\end{equation}
where $r^\star$ is the starting radius at which the fragmentation occurs and $\alpha = 2 - \beta$ (see Fig. \ref{global}a). For $r^\star = r_\circ$ we have $n_{cbm}^{act}=1$ particles/cm$^3$, as expected for a ``canonical GRB'' \citep{XIIBSGC} and in agreement with the apparent absence of a massive stellar wind in the CBM \citep{So06b,Fan06,Li07}.

The ${\cal R}$ parameter defined in Eq.(\ref{Rdef}) has to take into account both the effect of the fireshell fragmentation (${\cal R}_{shell}$) and of the effective CBM porosity (${\cal R}_{cbm}$):
\begin{equation}
{\cal R} \equiv {\cal R}_{shell} \times {\cal R}_{cbm}\, .
\label{rdef}
\end{equation}

The phenomenon of the clumpiness of the ejecta, whose measure is the filling factor, is an aspect well known in astrophysics.  For example, in the case of Novae the filling factor has been measured to be in the range $10^{-2}$--$10^{-5}$ \citep{Ede06}. Such a filling factor coincides, in our case, with ${\cal R}_{shell}$.

\section{Binaries as progenitors of GRB-SN systems}\label{binary}

The majority of the existing models in the literature appeal to a single astrophysical phenomenon to explain both the GRB and the SN \citep[``collapsar'', see e.g.][]{wb06}. On the contrary, a distinguishing feature of our theoretical approach is to differentiate between the SN and the GRB process. The GRB is assumed to occur during the formation process of a black hole. The SN is assumed to lead to the formation of a neutron star (NS) or to a complete disruptive explosion without remnants and, in no way, to the formation of a black hole. In the case of SN2006aj the formation of such a NS has been actually inferred by \citet{Mae06} because of the large amount of $^{58}$Ni ($0.05 M_\odot$). Moreover the significantly small initial mass of the SN progenitor star $M \approx 20 M_\odot$ is expected to form a NS rather than a black hole when its core collapses \citep{Mae06,Fe06,Maz06,No07}. In order to fulfill both the above requirement, we assume that the progenitor of the GRB and the SN consists of a binary system formed by a NS close to its critical mass collapsing to a black hole, and a companion star evolved out of the main sequence originating the SN. The temporal coincidence between the GRB and the SN phenomenon is explained in term of the concept of ``induced'' gravitational collapse \citep{rlet3,Mosca_Orale}. There is also the distinct possibility of observing the young born NS out of the SN \citep[see e.g.][and references therein]{Mosca_Orale}.

It has been often proposed that GRBs associated with SNe Ib/c, at smaller redshift $0.0085 < z < 0.168$ \citep[see e.g.][and references therein]{De06}, form a different class, less luminous and possibly much more numerous than the high luminosity GRBs at higher redshift \citep{Pi06,So04,Mae06,De06}. Therefore they have been proposed to originate from a separate class of progenitors \citep{Liang06,Cob06}. In our model this is explained by the nature of the progenitor system leading to the formation of the black hole with the smallest possible mass: the one formed by the collapse of a just overcritical NS \citep{RuffiniTF1,Mosca_Orale}.

The recent observation of GRB060614 at $z=0.125$ without an associated SN \citep{DV06,Mangano07} gives strong support to our scenario, alternative to the collapsar model. Also in this case the progenitor of the GRB appears to be a binary system composed of two NSs or a NS and a white dwarf (Caito et al., in preparation).

\section{Conclusions}

GRB060218 presents a variety of peculiarities, including its extremely large $T_{90}$ and its classification as an XRF. Nevertheless, a crucial point of our analysis is that we have successfully applied to this source our ``canonical GRB'' scenario.

Within our model there is no need for inserting GRB060218 in a new class of GRBs, such as the XRFs, alternative to the ``canonical'' ones. This same point recently received strong observational support in the case of GRB060218 \citep{la06} and a consensus by other models in the literature \citep{Kan06}.

The anomalously long $T_{90}$ led us to infer a monotonic decrease in the CBM effective density giving the first clear evidence for a fragmentation in the fireshell. This phenomenon appears to be essential in understanding the features of also other GRBs \citep[see e.g. GRB050315 in][and GRB970228 in Bernardini et al., in preparation]{Mosca_Orale}.

Our ``canonical GRB'' scenario originates from the gravitational collapse to a black hole and is now confirmed over a $10^6$ range in energy \citep[see e.g.][and references therein]{XIIBSGC}. It is clear that, although the process of gravitational collapse is unique, there is a large variety of progenitors which may lead to the formation of black holes, each one with precise signatures in the energetics. The low energetics of the class of GRBs associated with SNe, and the necessity of the occurrence of the SN, naturally leads in our model to identify their progenitors with the formation of the smallest possible black hole originating from a NS overcoming his critical mass in a binary system. For GRB060218 there is no need within our model for a new or unidentified source such as a magnetar or a collapsar.

GRB060218 is the first GRB associated with SN with complete coverage of data from the onset all the way up to $\sim 10^6$ s. This fact offers an unprecedented opportunity to verify theoretical models on such a GRB class. For example, GRB060218 fulfills the \cite{aa02} relation unlike other sources in its same class. This is particularly significant, since GRB060218 is the only source in such a class to have an excellent data coverage without gaps. We are currently examining if the missing data in the other sources of such a class may have a prominent role in their non-fulfillment of the \citet{aa02} relation \citep[Dainotti et al., in preparation; see also][]{ga06}.

\acknowledgements

We thank the Italian Swift Team (supported by ASI Grant I/R/039/04 and partly by the MIUR grant 2005025417) for the reduced Swift data, and Michael Kramer and Nino Panagia for the wording of the manuscript.


\begin{thebibliography}{99}

\bibitem[Amati et al.(2002)]{aa02}
Amati, L., Della Valle, M., Frontera, F., et al. 2002, \aap, 390, 81.

\bibitem[Barbier et al.(2006)]{Ba06}
Barbier, L., Barthelmy, S., Cummings, J., et al. 2006, GCN Circ. 4780.

\bibitem[Bernardini et al.(2006)]{030329}
Bernardini, M.G., Bianco, C.L., Chardonnet, P., et al. 2006, in ``Proceedings of the X$^{th}$ Marcel Grossmann Meeting'', World Scientific, 2459.

\bibitem[Bernardini et al.(2005)]{031203}
Bernardini, M.G., Bianco, C.L., Chardonnet, P., et al. 2005, \apjl, 634, L29.

\bibitem[Bianco \& Ruffini(2004)]{EQTS_ApJL}
Bianco, C.L., Ruffini, R. 2004, \apjl, 605, L1.

\bibitem[Bianco \& Ruffini(2005a)]{EQTS_ApJL2}
Bianco, C.L., Ruffini, R. 2005a, \apjl, 620, L23.

\bibitem[Bianco \& Ruffini(2005b)]{PowerLaws}
Bianco, C.L., Ruffini, R. 2005b, \apjl, 633, L13.

\bibitem[Campana et al.(2006)]{caa06}
Campana, S., Mangano, V., Blustin, A.J., et al. 2006, \nat, 442, 1008.

\bibitem[Cobb et al.(2006)]{Cob06}
Cobb, B.E., Bailyn, C.D., van Dokkum, P.G., Natarajan, P. 2006, \apjl, 645, L113.

\bibitem[Cusumano et al.(2006)]{ca06}
Cusumano, G., Barthelmy, S., Gehrels, N., et al. 2006, GCN Circ. 4775.

\bibitem[Della Valle(2006)]{De06}
Della Valle, M. 2006, AIP. Con.Proc. 836, 367.

\bibitem[Della Valle et al.(2006)]{DV06}
Della Valle, M., Chincarini, G., Panagia, N., et al. 2006, \nat, 444, 1050.

\bibitem[Ederoclite et al.(2006)]{Ede06}
Ederoclite, A., Mason, E., Della Valle, M., et al. 2006, \aap, 459, 875.

\bibitem[Fatkhullin et al.(2006)]{fa06}
Fatkhullin, T.A., Vlasyuk, V.V., Sokolov, V.V., et al. 2006, GCN Circ. 4809.

\bibitem[Fan et al.(2006)]{Fan06}
Fan, Y., Piran, T., Xu, D. 2006, JCAP 0609, 013.

\bibitem[Ferrero et al.(2006)]{Fe06}
Ferrero, P., Palazzi, E., Pian, E., Savaglio, S., in proceedings of ``The Multicoloured Landscape of Compact Objects and their Explosive Progenitors: Theory vs Observations'', astro-ph/0610417.

\bibitem[Ghisellini et al.(2006)]{ga06}
Ghisellini, G., Ghirlanda, G., Mereghetti, S., et al. 2006, \mnras, 372, 1699.

\bibitem[Guetta \& Della Valle(2007)]{Gue06}
Guetta, D., Della Valle, M. 2007, \apjl, 657, L73.

\bibitem[Kaneko et al.(2006)]{Kan06} 
Kaneko, Y., Ramirez-Ruiz, E., Granot, J., et al. 2006 \apj, 654, 385.

\bibitem[Kennea et al.(2006)]{Ka06}
Kennea, J., Burrows, D., Cusumano, G., Tagliaferri, G. 2006, GCN Circ. 4776.

\bibitem[Liang et al.(2006a)]{la06}
Liang, E.W., Zhang, B.-B., Stamatikos, M., et al. 2006a, \apjl, 653, L81.

\bibitem[Liang et al.(2006b)]{Liang06}
Liang, E., Zhang, B., Virgili F., Dai, Z.G., astro-ph/0605200

\bibitem[Li(2007)]{Li07}
Li, L.-X. 2007, \mnras, 375, 240.

\bibitem[Mangano et al.(2007)]{Mangano07}
Mangano, V. Holland, S.T., Malesani, D., et al. 2007, arXiv:0704.2235.

\bibitem[Masetti et al.(2006)]{ma06}
Masetti, N., Palazzi, E., Pian, E., Patat, F. 2006, GCN Circ. 4803.

\bibitem[Maeda et al.(2007)]{Mae06}
Maeda, K., Kawabata, K., Tanaka, M., et al. 2007, \apjl, 658, L5.

\bibitem[Mazzali et al.(2006)]{Maz06}
Mazzali, P., Deng, J., Nomoto, K., et al. 2006, \nat, 442, 1018.

\bibitem[Mirabal et al.(2006)]{Mi06}
Mirabal, N., Halpern, J., Thorstensen, J., Terndrup, D. 2006, \apjl, 643, L99.

\bibitem[Modjaz et al.(2006)]{Modjaz06}
Modjaz, M., Stanek, K.Z., Garnavich, P.M., et al. 2006, \apjl, 645, L21.

\bibitem[Nomoto et al.(2007)]{No07} 
Nomoto, K., Tominaga, N., Tanaka, M., et al. 2007, astro-ph/0702472.

\bibitem[Pe'er et al.(2007)]{pa07}
Pe'er, A., Ryde, F., Wijers, R.A.M.J., et al. 2007, astro-ph/0703734.

\bibitem[Pian et al.(2006)]{Pi06}
Pian, E., Mazzali, P., Masetti, N., et al. 2006, \nat, 442, 1011.

\bibitem[Piran(2004)]{p04}
Piran, T. 2004, Rev.Mod.Phys., 76, 1143.

\bibitem[Ruffini(2006)]{RuffiniTF1}
Ruffini, R. 2006, in the ``Proceedings of the XI$^{th}$ Marcel Grossmann meeting'', World Scientific, in press.

\bibitem[Ruffini et al.(2005a)]{rubr2}
Ruffini, R., Bernardini, M.G., Bianco, C.L., et al. 2005a, AIP Con.Proc. 782, 42.

\bibitem[Ruffini et al.(2006)]{050315}
Ruffini, R., Bernardini, M.G., Bianco, C.L., et al. 2006, \apj, 645, L109.

\bibitem[Ruffini et al.(2007a)]{XIIBSGC}
Ruffini, R., Bernardini, M.G., Bianco, C.L., et al. 2007a, AIP Con.Proc. 910, 55.

\bibitem[Ruffini et al.(2007b)]{Mosca_Orale}
Ruffini, R., Bernardini, M.G., Bianco, C.L., et al. 2007b, arXiv:0705.2456.

\bibitem[Ruffini et al.(2007c)]{Venezia_Orale}
Ruffini, R., Bernardini, M.G., Bianco, C.L., et al. 2007c, arXiv:0705.2453.

\bibitem[Ruffini et al.(2001a)]{rlet1}
Ruffini, R., Bianco, C.L., Chardonnet, P., et al. 2001a, \apjl, 555, L107.

\bibitem[Ruffini et al.(2001b)]{rlet2}
Ruffini, R., Bianco, C.L., Chardonnet, P., et al. 2001b, \apjl, 555, L113.

\bibitem[Ruffini et al.(2001c)]{rlet3}
Ruffini, R., Bianco, C.L., Chardonnet, P., et al. 2001c, \apjl, 555, L117.

\bibitem[Ruffini et al.(2003)]{rubr}
Ruffini, R., Bianco, C.L., Chardonnet, P., et al. 2003, AIP Con.Proc. 668, 16.

\bibitem[Ruffini et al.(2004)]{spectr1}
Ruffini, R., Bianco, C.L., Chardonnet, P., et al. 2004, Int.J.Mod.Phys.D, 13, 843.

\bibitem[Ruffini et al.(2005b)]{fil}
Ruffini, R., Bianco, C.L., Chardonnet, P., et al. 2005b, Int.J.Mod.Phys.D, 14, 97.

\bibitem[Ruffini et al.(2000)]{rswx00} 
Ruffini, R., Salmonson, J.D., Wilson, J.R., Xue, S.S. 2000, \aap, 359, 855.

\bibitem[Sakamoto et al.(2006)]{Sa06}
Sakamoto, T., Barbier, L., Barthelmy, S., et al. 2006, GCN Circ. 4822.

\bibitem[Soderberg et al.(2004)]{So04}
Soderberg, A.M., Kulkarni, S.R., Berger, E., et al. 2004, \nat, 430, 648.

\bibitem[Soderberg et al.(2006a)]{So06a}
Soderberg, A.M., Berger, E., Schmidt, B.P., et al. 2006a, GCN Circ. 4804.

\bibitem[Soderberg et al.(2006b)]{So06b}
Soderberg, A.M., Kulkarni, S.R., Nakar, E., et al. 2006b, \nat, 442, 1014.

\bibitem[Sollerman et al.(2006)]{Sollermann06}
Sollerman, J., Jaunsen, A.O., Fynbo, J.P.U., et al. 2006, \aap, 454, 503.

\bibitem[Toma et al.(2006)]{Toma06}
Toma, K., Ioka, K., Sakamoto, T., Nakamura, T. 2007, \apj, 659, 1420.

\bibitem[Wiersema et al.(2007)]{Wi07}
Wiersema, K., Savaglio, S., Vreeswijk, P.M., et al. 2007 \aap, 464, 529.

\bibitem[Woosley \& Bloom(2006)]{wb06}
Woosley, S.E., Bloom, J.S. 2006, \araa, 44, 507.

\end{thebibliography}
\end{document}